\documentclass[aps,prb,amsmath,amssymb,twocolumn,10pt,footinbib]{revtex4-1}
\usepackage{graphicx}
\usepackage{hyperref}
\usepackage{dcolumn}
\usepackage{bm}

\begin{document}
\title{Electronic states induced by nonmagnetic defects in two-dimensional topological insulators}

\author{Vladimir~A.~Sablikov and Aleksei~A.~Sukhanov}

\affiliation{V.A. Kotel’nikov Institute of Radio Engineering and Electronics, Russian Academy of Sciences,
Fryazino, Moscow District, 141190, Russia}

\begin{abstract}
We study in-gap electronic states induced by a nonmagnetic defect with short-range potential in two-dimensional topological insulators and trace their evolution as the distance between the defect and the boundary changes. The defect located far from the boundary is found to produce two bound states independently of the sign of its potential. The states are classified as electronlike and holelike. Each of these states can have two types of the spatial distribution of the electron density. The first-type states have a maximum of the density in the center and the second-type ones have a minimum. When the defect is coupled with the boundary, the bound states are transformed correspondingly into resonances of two types and take up the form of the edge states flowing around the defect. Under certain conditions, two resonances interfere giving rise to the formation of a bound state embedded into the continuum spectrum of the edge states flowing around the defect. We calculate the spatial distribution of the electron density in the edge states flowing around the defect and estimate the charge accumulated near the defect. The current density field of the edge states flowing around the defect contains two components one of which flows around the defect and the other circulates around it.
\end{abstract}
\maketitle

\section{Introduction}
\label{Intro}
The presence of gapless edge states at the interface of topologically non-equivalent crystals is a hallmark of two-dimensional (2D) topological insulators (TIs)~\cite{Kane}. In these states the electrons move along the boundary and their spin is locked to the momentum because of strong spin-orbit interaction. Such helical edge states are protected against scattering by weak non-magnetic impurities and disorders. Nevertheless, experiments reveal a noticeable backscattering of electrons~\cite{Konig1,Konig2,Roth,Gusev,Konig3,Nowack,Spanton}, the mechanism of which is not yet known~\cite{Vayrynen,Vayrynen1}. 

Backscattering of electrons in the edge states can occur as a result of an inelastic process due to electron-electron interactions and the presence of a defect potential~\cite{Schmidt}. The effect of the electron-electron interaction in the vicinity of the defect essentially depends on the charge and spin structure of the electron cloud which forms near it. In this regard, of great importance is the question about the electronic states induced by impurities and other structural imperfections, especially in the case where the defect is located near the boundary. One-dimensional models of coupling between the edge states and the defect turn out to be insufficient to describe the experiments~\cite{Xu,Wu,Schmidt}.

Electronic states induced by a defect were studied for three dimensional (3D) TIs where the defect is located on the surface. In this case, the electron cloud around the defect is formed by 2D electronic states propagating along the surface. Their interference leads to a variety of the electron density configurations~\cite{Zhou,Guo,Wang,Biswas} and even to the changes in the surface state spectra~\cite{Black-Schaffer1,Black-Schaffer2}.

In 2D TIs, the electron cloud around a defect also exists but its structure is substantially different since the electron density configuration is formed mainly by evanescent modes decaying in the plane. It is essential that the electron cloud can not be described within a one-dimensional (1D) model. Electronic structures formed in this case are currently poorly understood.

Defect-induced electronic states in 2D TIs were studied mostly in the case where the defect is located deep in the bulk and decoupled from the boundary. In Ref.~\onlinecite{Shan}, the defect was considered as a hole, at the edges of which the wave function is zero. In this case, the bound states are in essence the edge states circulating around the hole with quantized angular momentum. Although this model captures some properties of the defect-induced states, it is far from reality. Under realistic conditions the wave function is not zero in the defect. The bound states appearing in the Gaussian potential were investigated numerically for a number of material parameters~\cite{J_Lu,Shen}, but no general conclusions were made about their spectra, the electronic structure, and the conditions under which they exist.

Defects interacting with the boundaries were studied in the case of a slab of 2D TI. In this case, the defect is coupled with two boundaries. Numerical calculations with using Green's function method combined with tight-binding approach~\cite{Lee} have shown that the bound state spectrum differs from that in the continuous model. Particularly, it contains two states bound on one defect with short range potential rather than one state as in the continuous model~\cite{J_Lu}.

In recent work~\cite{Sablikov} we investigated analytically the bound states induced by a non-magnetic defect in the bulk of 2D TIs for defects with short-range potential. It turned out that the defect creates two bound states which are classified as electronlike and holelike. This is in contrast to the defects in topologically trivial insulators where only one bound state exists in a short-range potential. The bound states exist for both positive and negative potentials. In turn these states can be also of two types depending on whether the electron density has a maximum or a minimum in the point of the defect location. Another interesting feature of the 2D TIs is an unusual dependence of the bound-state energies on the defect potential. As the potential increases, the energies of both electronlike and holelike states tend correspondingly to two different limiting values, which lie within the gap.

In this paper we address the general problem of a defect coupled with the edge states in 2D TIs. We clarify how the bulk bound states are modified with decreasing the distance between the defect and the boundary and how the edge states are distorted by the defect. It turns out that the edge states and the bulk bound states transform into a set of eigenstates which have the form of the edge states flowing around the defect. These states have resonances of the electron density in the vicinity of the defect when the energy is close to the energy of the bulk bound states. Correspondingly, there are two types of the resonances. 

Under certain conditions two resonances of different types can interfere with each other giving rise to the formation of a bound state with localized wave function in the continuum of the edge states. 

We study the spatial distribution of the electron density and current density in the states flowing around the defect and estimate the charge accumulated near the defect at a given Fermi energy.

The outline of the paper is as follows. In Section~\ref{bound_flowing_states} we present analytical calculations showing the presence of the states flowing around the defect and the bound states in the continuum. Section~\ref{bound_states_bulk} gives the detailed results for the bound states in the bulk of 2D TI. Section~\ref{states_near_boundary} deals with the electron states in the case where the defect is located at a finite distance from the boundary. In Sec.~\ref{e_dens}, we study the electron density distributions for resonant states, estimate the excess electron density accumulated near the defect and consider the patterns of the electron current. We finish the paper with a discussion and conclusions in Sec.~\ref{discuss}.

\section{Bound states and edge states flowing around the defect}\label{bound_flowing_states}
Our study is based on the model of the 2D TIs proposed by Bernevig, Hughes, and Zhang (BHZ) for HgTe/CdTe quantum wells~\cite{BHZ}. The 2D TI is described by the Hamiltonian
\begin{equation}
 H_0=
\begin{pmatrix}
 h(\mathbf{k}) & 0\\
 0 & h^*(-\mathbf{k})
\end{pmatrix}\,,
\label{Hamiltonian0}
\end{equation} 
where $\mathbf k$ is momentum operator and
\begin{equation}
h(\mathbf{k})=
\begin{pmatrix}
 M\!-\!(B\!+\!D)k^2 & A(k_x+ik_y)\\
 A(k_x-ik_y) & -M\!+\!(B\!-\!D)k^2
\end{pmatrix}\,,
\label{h-Hamiltonian}
\end{equation} 
with $M$, $A$, $B$ and $D$ being the model parameters. The topological phase is realized when $MB>0$.~\cite{BHZ,Lu} In the case of the HgTe/CdTe wells, the parameters $M, B, D<0$, and $A>0$. The basis set of wave functions is \mbox{$\{|E_1\uparrow\rangle,|H_1\uparrow\rangle,|E_1\downarrow\rangle,|H_1\downarrow\rangle\}$} where $|E_1\uparrow\rangle$ and $|E_1\downarrow\rangle$ are superpositions of the electron states of $s$-type and light-hole states of  $p$-type with spin up and spin down; $|H_1\uparrow\rangle$ and $|H_1\downarrow\rangle$ are the heavy-hole $p$-type states with opposite spins. In what follows we will restrict ourselves by considering the symmetric model where $D=0$.

\begin{figure}
\centerline{\includegraphics[width=.8\linewidth]{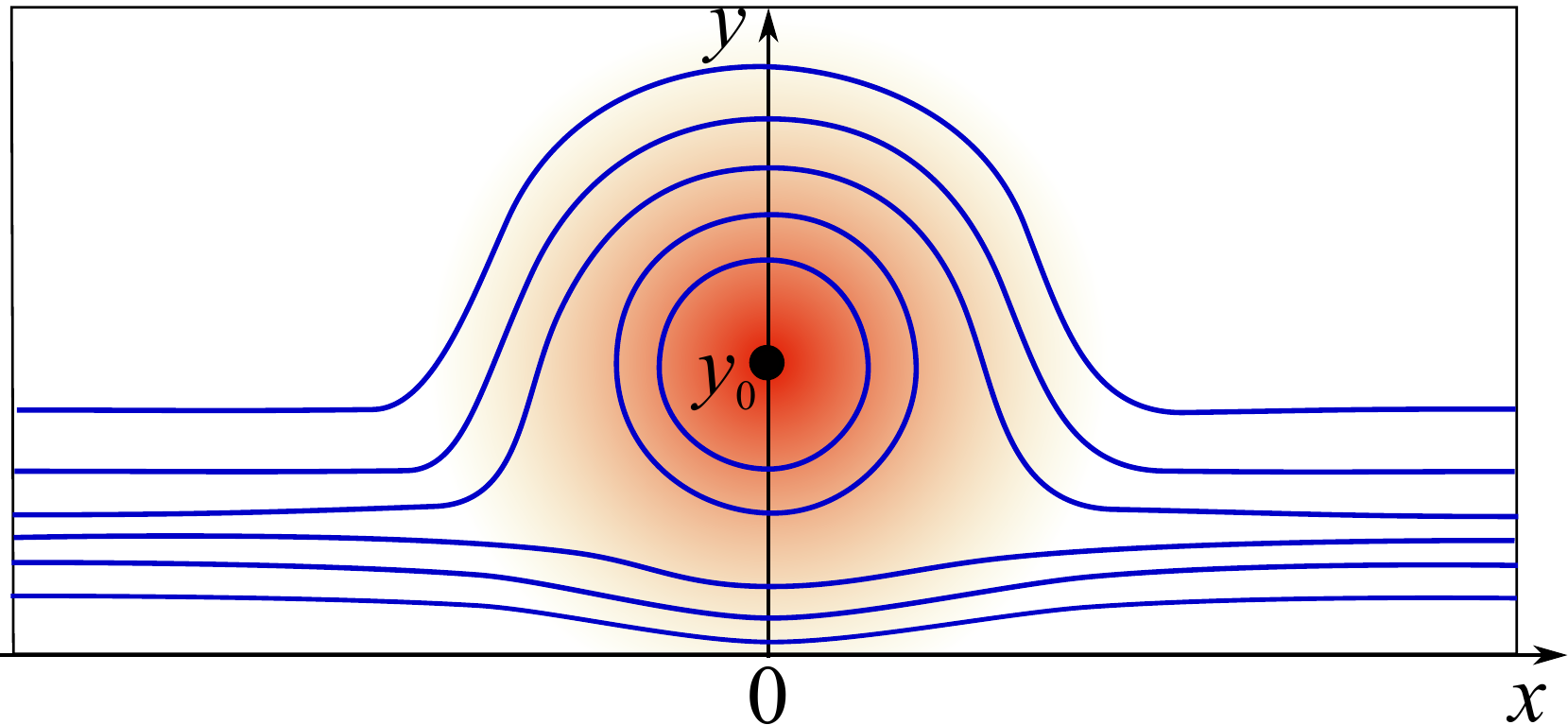}}
\caption{(Color online) A schematic view of a defect located at the distance $y_0$ from the boundary of the 2D TI. The darkened area shows the particle cloud near the defect. Lines represent the particle flows.}
\label{schematic}
\end{figure}

Let us use the Cartesian coordinates, with the $x$ axis coinciding with the boundary (Fig.~\ref{schematic}). The TI lies at $y>0$ and the defect is located in the point $x=0, y=y_0$. We consider the defect described by a potential $V(x,y-y_0)$ localized in a small region. Since the defect is non-magnetic, the total Hamiltonian $H_0+V(x,y-y_0)$ is separated into spin blocks. For spin-up electrons, the Schr\"odinger equation reads as
\begin{equation}
\left[E\sigma_0-h(\mathbf{k})\right]\Psi(x,y)=\sigma_0 V(x,y-y_0)\Psi(x,y)\,,
\label{Schrodinger_eq}
\end{equation} 
where $\sigma_0$ is a $2\times2$ unit matrix, $\Psi(x,y)$ is a spinor $\left(\psi_1(x,y),\psi_2(x,y)\right)^T$. The wave functions are supposed to vanish at $y\to \infty$ and equal zero at $y=0$.

In what follows we will use dimensionless variables
\begin{equation}
 \begin{array}{l}
 \varepsilon\!=\!E/|M|,\; \{x',y'\}\!=\!\{x,y\}\sqrt{M/B},\; a\!=\!A/\sqrt{MB},\\
 v(x',y')=V(x,y)/|B|,\;\,\, b=y_0 \sqrt{M/B},
 \end{array}
\label{dim_less_param}
\end{equation} 
and for convenience will omit the prime in the variables $x', y'$.

The 2D problem~(\ref{Schrodinger_eq}) can be solved by using the Fourier and Laplace transforms over $x$ and $y$:
\begin{align}
 \widetilde{\Psi}(k,p)=&\int_{-\infty}^{\infty}\!dx e^{-ikx}\int_{0}^{\infty}\!dye^{-py}\Psi(x,y)\,,\\
 \Psi(x,y)=&\int\limits_{-\infty}^{\infty}\!\frac{dk}{2\pi} e^{ikx}\int\limits_{c-i\infty}^{c+i\infty}\!\frac{dp}{2\pi i}e^{py}\widetilde{\Psi}(k,p)\,.
 \label{F-L_Psi}
\end{align}
When applying this transformation to Eq.~(\ref{Schrodinger_eq}) one needs to calculate the Fourier and Laplace transforms of the product $v(x,y)\Psi(x,y)$. We suppose that the region, where the defect potential is localized, is small compared with the characteristic length scale of the wave function. In this case the integral can be approximated as
\begin{equation}
 \int\limits_{-\infty}^{\infty}\!dx e^{-ikx}\!\int\limits_0^{\infty}\!dy e^{-py}v(x,y)\Psi(x,y)\!\approx \!\widetilde{v}(k,p)e^{-bp}\overline{\Psi},
\end{equation} 
where $\overline{\Psi}=\Psi(x=0,y=b)$ is the wave function at the defect position and $\widetilde{v}(k,p)$ is the Fourier and Laplace transforms of $v(x,y)$. In such a way we arrive at the following equation:
\begin{equation}
 [\varepsilon-h(k,p)]\widetilde{\Psi}(k,p)=\sigma_z\Phi(k)+\sigma_0 \widetilde{v}(k,p)e^{-bp}\overline{\Psi}\,.
\label{Fourier-Laplace1}
\end{equation} 
Here, 
$[\varepsilon-h(k,p)]$ is the matrix with elements $a_{ij}(\varepsilon,k,p)$:
\begin{equation}
 \begin{array}{ll}
 a_{11}=\varepsilon+1-k^2+p^2, & a_{12}=-a(k+p),\\
 a_{21}=-a(k-p), & a_{22}=\varepsilon-1+k^2-p^2,
 \end{array}
\end{equation} 
$\sigma_z$ is the Pauli matrix, $\Phi(k)$ is the Fourier transform of the normal derivative of $\Psi(x,y)$ at the boundary: 
\begin{equation}
 \Phi(k)=\int_{-\infty}^{\infty}dx e^{-ikx}\frac{\partial\Psi(x,y)}{\partial y}\bigg|_{y=0}.
\end{equation} 

We are going to get a system of linear equations for the components of the spinor $\overline{\Psi}$ which will allow one to determine the eigenenergy spectrum. This idea is implemented as follows.

Solving Eq.~(\ref{Fourier-Laplace1}) with respect to $\widetilde{\Psi}(k,p)$ and using Eq.~(\ref{F-L_Psi}), we obtain the following expression for the wave function:
\begin{multline}
 \Psi(x,y)=\int\limits_{-\infty}^{\infty}\frac{dk}{2\pi}e^{ikx}\int\limits_{c-i\infty}^{c+i\infty}\frac{dp}{2\pi i}\frac{e^{py}} {\Delta(\varepsilon,k,p)}\\
 \times\left[D_0(\varepsilon,k,p)\Phi(k)+v(k,p)e^{-bp}D_1(\varepsilon,k,p)\overline{\Psi}\right],
\label{Psi_Phi_Psi-bar}
\end{multline}
where $\Delta(\varepsilon,k,p)$ is the determinant of the matrix in the left-hand side of Eq.~(\ref{Fourier-Laplace1}) which has the form:
\begin{equation}
 \Delta(\varepsilon,k,p)=-\left[p^2-p_1^2(\varepsilon,k)\right]\left[p^2-p_2^2(\varepsilon,k)\right],
\end{equation} 
with
\begin{equation}
 p_{1,2}(\varepsilon,k)=\sqrt{k^2+a^2/2-1\pm\sqrt{a^2(a^2-4)/4+\varepsilon^2}}
 \label{p12}
\end{equation} 
and $\mathrm{Re}p_{1,2}(\varepsilon,k)\ge 0$; $D_0(\varepsilon,k,p)$ and $D_1(\varepsilon,k,p)$ are the following matrices:
\begin{equation}
 D_0=
 \begin{pmatrix}
  a_{22}(\varepsilon,k,p) & a_{12}(\varepsilon,k,p)\\
  -a_{21}(\varepsilon,k,p) & -a_{11}(\varepsilon,k,p)
 \end{pmatrix}\,,
\end{equation}
\begin{equation}
D_1=
 \begin{pmatrix}
  a_{22}(\varepsilon,k,p) & -a_{12}(\varepsilon,k,p)\\
  -a_{21}(\varepsilon,k,p) & a_{11}(\varepsilon,k,p)
 \end{pmatrix}.
\end{equation} 

Let us now turn to the requirement that $\Psi(x,y)$ should not diverge in the limit $y\to \infty$. Equation~(\ref{Psi_Phi_Psi-bar}) shows that $\Psi(x,y\to \infty)\nrightarrow \infty$ when the expression in the square brackets equals zero at $p=p_{1,2}(\varepsilon,k)$. This gives us two equations that relate $\Phi(k)$ and $\overline{\Psi}$. Correspondingly, there are four equations for their spinor components. However, one can show that only two of these equations are independent because the matrix elements $a_{ij}(\varepsilon,k,p)$ at $p=p_1$ and $p=p_2$ are connected by joint equation $\Delta(\varepsilon,k,p_{1,2})=0$. In such a way we arrive at the following equation:
\begin{equation}
 A(\varepsilon,k)\Phi(k)+B(\varepsilon,k)\overline{\Psi}=0,
\label{Phi-Psi-bar}
\end{equation} 
where $A(\varepsilon,k)$ and $B(\varepsilon,k)$ are matrices
\begin{equation}
 A(\varepsilon,k)=
 \begin{pmatrix}
  a_{22}(\varepsilon,k,p_1) & a_{12}(\varepsilon,k,p_1)\\
  a_{22}(\varepsilon,k,p_2) & a_{12}(\varepsilon,k,p_2)
 \end{pmatrix}\,,
\end{equation} 
\begin{multline}
 B(\varepsilon,k)=\\
 \begin{pmatrix}
  v(k,p_1)a_{22}(\varepsilon,k,p_1)e^{-bp_1} & -v(k,p_1)a_{12}(\varepsilon,k,p_1)e^{-bp_1}\\
  v(k,p_2)a_{22}(\varepsilon,k,p_2)e^{-bp_2} & -v(k,p_2)a_{12}(\varepsilon,k,p_2)e^{-bp_2}
 \end{pmatrix}.
 \label{B}
\end{multline}

Solving Eq.~(\ref{Phi-Psi-bar}) with respect to $\Phi(k)$ we obtain an explicit expression for $\Phi(k)$:
\begin{equation}
 \Phi(k)=-\frac{A'(\varepsilon,k)B(\varepsilon,k)}{\Delta_1(\varepsilon,k)}\overline{\Psi}+C(\varepsilon)\chi(\varepsilon,k)\delta[k-k_0(\varepsilon)],
\label{Phi-Psi_1}
\end{equation} 
where $\Delta_1(\varepsilon,k)$ is the determinant of the matrix $A(\varepsilon,k)$ and $A'(\varepsilon,k)$ is the following matrix:
\begin{equation}
 A'(\varepsilon,k)=
 \begin{pmatrix}
  a_{12}(\varepsilon,k,p_2) & -a_{12}(\varepsilon,k,p_1)\\
  -a_{22}(\varepsilon,k,p_2) & a_{22}(\varepsilon,k,p_1)
 \end{pmatrix}\,.
\label{A}
\end{equation} 
The second term in Eq.~(\ref{Phi-Psi_1}) arises because of the singularity of the matrix $A(\varepsilon,k)$ in accordance with the general theory of singular matrices~\cite{Ben-Israel}. It describes the contribution of the edge states in the pure TI into the electronic states formed in the presence of the defect. $k_0(\varepsilon)$ is a root of the determinant $\Delta_1(\varepsilon,k)$ which gives exactly the spectrum of the edge states in the absence of the defect:
\begin{equation}
 k_0(\varepsilon)=-\frac{\varepsilon}{a}\,.
\label{edge_spectrum}
\end{equation} 
Further in Eq.~(\ref{Phi-Psi_1}), the coefficient $C(\varepsilon)$ is a normalization constant, $\chi(\varepsilon,k)$ is a spinor which is expressed via the matrix elements $a_{ij}(\varepsilon,k,p)$ at $p=p_{1,2}$. Using Eqs~(\ref{p12}) and (\ref{edge_spectrum}) it is easy to show that $\chi(\varepsilon,k)$ coincides with the spinor of the edge states:
\begin{equation}
 \chi=\binom{~1}{-1}.
\end{equation} 

Let us now apply Eq.~(\ref{Psi_Phi_Psi-bar}) to calculate $\overline{\Psi}$. To this end, we set $x=0$ and $y=b$ and exclude $\Phi(k)$ using Eq.~(\ref{Phi-Psi_1}). Finally, we obtain the equation which determines $\overline{\Psi}$: 
\begin{equation}
 \left(\sigma_0-\mathcal{K}(\varepsilon)\right)\overline{\Psi}=C(\varepsilon)\mathcal{F}(\varepsilon)\chi,
\label{Psi-bar}
\end{equation} 
where $\mathcal{K}(\varepsilon)$ and $\mathcal{F}(\varepsilon)$ are the following matrices
\begin{multline}
 \mathcal{K}(\varepsilon)=\int\limits_{-\infty}^{\infty}\frac{dk}{2\pi}\left[\frac{1}{4a_{\varepsilon}\Delta_1(\varepsilon,k)}\mathcal{D}_0(\varepsilon,k)A'(\varepsilon,k)B(\varepsilon,k)\right.\\
 +\biggl.\int\limits_{-i\infty}^{i\infty}\frac{dp}{2\pi i}\frac{v(k,p)}{\Delta(\varepsilon,k,p)}D_1(\varepsilon,k,p)\biggr],
\label{K}
\end{multline}
\begin{equation}
 \mathcal{F}(\varepsilon)=\frac{1}{4a_{\varepsilon}}\mathcal{D}_0\bigl(\varepsilon,k_0(\varepsilon)\bigr).
\end{equation}
Here $a_{\varepsilon}=\sqrt{a^2(a^2/4-1)+\varepsilon^2}$ and $\mathcal{D}_0(\varepsilon,k)$ denotes the matrix
\begin{equation}
 \mathcal{D}_0(\varepsilon,k)\!=\!\frac{e^{-bp_1}}{p_1}D_0(\varepsilon,k,\!-p_1)\!-\!\frac{e^{-bp_2}}{p_2}D_0(\varepsilon,k,\!-p_2).
\end{equation} 

Equation~(\ref{Psi-bar}) has solutions of two kinds depending on the determinant of the matrix $\left(\sigma_0-\mathcal{K}(\varepsilon)\right)$ 
\begin{equation}
 \Delta_{\Psi}(\varepsilon)=\bigl(1-\mathcal{K}_{11}(\varepsilon)\bigr)\bigl(1-\mathcal{K}_{22}(\varepsilon)\bigr)-\mathcal{K}_{12}(\varepsilon)\mathcal{K}_{21}(\varepsilon)
\end{equation} 

First, if $\Delta_{\Psi}(\varepsilon)\ne 0$, the root of Eq.~(\ref{Psi-bar}) is
\begin{equation}
 \overline{\Psi}(\varepsilon)=\frac{C(\varepsilon)}{\Delta_{\Psi}(\varepsilon)}\left[\sigma_0-\mathcal{K}'(\varepsilon)\right]\mathcal{F}(\varepsilon)\chi,
\label{Psi-bar_flowing}
\end{equation} 
where
\begin{equation}
 \sigma_0-\mathcal{K}'(\varepsilon)=
 \begin{pmatrix}
  1-\mathcal{K}_{22}(\varepsilon) & \mathcal{K}_{12}(\varepsilon)\\
  \mathcal{K}_{21}(\varepsilon) & 1-\mathcal{K}_{11}(\varepsilon)
 \end{pmatrix}\,.
\end{equation} 

An alternative is the case where 
\begin{equation}
\Delta_{\Psi}(\varepsilon)=0\,.
\label{epsilon_0}
\end{equation} 
Let $\varepsilon_0$ is a root of this equation. When $\varepsilon=\varepsilon_0$, Eq.~(\ref{Psi-bar}) has a solution if $C=0$. This solution reads as
\begin{equation}
 \overline{\Psi}(\varepsilon)=C_{bs}\binom{1}{(1-\mathcal{K}_{11})\bigm/\mathcal{K}_{12}}\Biggm|_{\varepsilon=\varepsilon_0}\,,
\label{Psi-bar_BS}
\end{equation} 
with the constant $C_{bs}$ being determined by the normalization. 

It is worth noting that in the first case, $C(\varepsilon)$ should turn to zero when $\varepsilon$ tends to $\varepsilon_0$. Otherwise, $\Psi(x,y)$ will not be normalized. Thus $\overline{\Psi}(\varepsilon)$ is determined by Eq.~(\ref{Psi-bar_flowing}) for any $\varepsilon\ne \varepsilon_0$.  But if $C(\varepsilon)$ is exactly zero, the solution $\overline{\Psi}$ is given by Eq.~(\ref{Psi-bar_BS}).

In order to clarify the nature of these solutions we consider the asymptotics of $\Psi(x,y)$ at $x\to \pm\infty$. The asymptotic behavior of $\Psi(x,y)$ is easily found from Eqs~(\ref{Psi_Phi_Psi-bar}) and (\ref{Phi-Psi_1}). It has the following form:  
\begin{multline}
 \Psi(x\to\infty,y)\simeq\frac{i e^{ikx}}{8a_{\varepsilon}\frac{\partial \Delta_1}{\partial k}}\left[\frac{e^{-p_2y}}{p_2}D_0(\varepsilon,k,-p_2)\right.\\ 
 -\left.\frac{e^{-p_1y}}{p_1}D_0(\varepsilon,k,-p_1)\right]A'(\varepsilon,k)B(\varepsilon,k)\overline{\Psi}(\varepsilon)\Biggm|_{k=k_0(\varepsilon)}.
 \label{Psi_asymptotic}
\end{multline}

In the case where $\overline{\Psi}(\varepsilon)$ is determined by Eq.~(\ref{Psi-bar_flowing}), one can show that $\Psi(x\to\infty,y)$ never equals zero and is proportional to $\exp[ik_0x]$.  Hence, these states propagate along the edge and flow around the defect.  We will call them the edge states flowing around the defect. They have the continuous spectrum defined by Eq.~(\ref{edge_spectrum}), which coincides with the spectrum of the edge states without defects. The constant $C(\varepsilon)$ can be found by appropriate normalization.

At a discrete energy $\varepsilon=\varepsilon_0$ the wave function should be square integrable and hence the amplitude in its asymptotics, given by Eq.~(\ref{Psi_asymptotic}), should be zero. Using the specific expressions for matrices $A'(\varepsilon,k)$ and $B(\varepsilon,k)$, given by Eqs.~(\ref{A}) and (\ref{B}), one can easily show that $A'(\varepsilon,k)B(\varepsilon,k)\overline{\Psi}=0$ if 
\begin{equation}
 \overline{\Psi}=\overline{\psi}\binom{1}{1}.
 \label{Psi-bar_BS1}
\end{equation} 
Thus, when $\overline{\Psi}(\varepsilon_0)$ satisfies Eq.~(\ref{Psi-bar_BS1}), a bound state can arise in the continuum of edge states. Comparing Eq.~(\ref{Psi-bar_BS}), which defines $\overline{\Psi}(\varepsilon)$, and Eq.~(\ref{Psi-bar_BS1}) we arrive at the following equation for the elements of the $\mathcal{K}$ matrix:
\begin{equation}
 1-\mathcal{K}_{11}(\varepsilon)-\mathcal{K}_{12}(\varepsilon)=0\,.
 \label{1-K11-K12}
\end{equation} 

Importantly, this equation must be satisfied together with Eq.~(\ref{epsilon_0}) that gives the necessary condition for the bound state to exist. At this point, one should take into account that the elements of the $\mathcal{K}$ matrix depend not only on the energy $\varepsilon$, but also on the defect potential $v(x,y)$. Therefore, the system of Eqs.~(\ref{epsilon_0}) and (\ref{1-K11-K12}) determines the energy $\varepsilon_{bs}$ of the bound state in the continuum and the defect potential $v_{bs}$ at which this state arises.

Following, we present the results of specific calculations of the bound states and the states flowing around the defect.

\section{Bound states in the bulk}\label{bound_states_bulk}
We start by considering the limit of $b\to \infty$, which describes the bound states for a defect located in the bulk. 
When $b\to \infty$, the right-hand side of Eq.~(\ref{Psi-bar}) goes to zero and the nondiagonal components of the $\mathcal{K}(\varepsilon)$ matrix defined in Eq.~(\ref{K}) also vanish. As a result, Eq.~(\ref{Psi-bar}) decouples into two independent homogeneous equations for the components of the spinor $\overline{\Psi}=(\overline{\psi}_1,\overline{\psi}_2)^T$. Correspondingly, there are two kinds of bound states with different pseudospin components of the wave function at the defect.

There is a solution in which $\overline{\psi}_1\ne 0$ and $\overline{\psi}_2=0$. Since $\psi_1$ corresponds to the $|E1\rangle$ component of the basis set of wave functions, the states of this kind can be conventionally called the electronlike states. In another solution, in contrast $\overline{\psi}_1=0$ and $\overline{\psi}_2 \ne 0$. We call them the holelike states. 

The eigenenergies of the states of both species are determined by Eq.~(\ref{epsilon_0}). In the limit $b\to \infty$, Eq.~(\ref{epsilon_0}) decouples into two equations. Correspondingly, there are two solutions for electron-like and hole-like states: $\varepsilon_e$ and $\varepsilon_h$. The bound-state energies depend on the defect potential $v(x,y)=vf(x,y)$. Although the particle-hole symmetry is broken due to the defect potential, the following symmetry relation holds for the energies of the electron-like and hole-like bound states: 
\begin{equation}
\varepsilon_e(v)=-\varepsilon_h(-v).                                                                                                                                                                                                                                                                                                                                                                                                                                                                                        
\label{e-h_symmetry}
\end{equation}  

Specific calculations of the bound-state energies and the electron density were carried out for the defect potential of two forms: the Gaussian function $v(x,y)=v\Lambda^2/\pi\exp[-\Lambda^2(x^2+y^2)]$ with the characteristic radius $\Lambda^{-1}$, and the $v(x,y)=v/\pi\delta\left(x^2+y^2\right)$ with regularizing cut-off at $\Lambda$ when integrating over the wave vector. Both cases give similar results.

\begin{figure}
\centerline{\includegraphics[width=.9\linewidth]{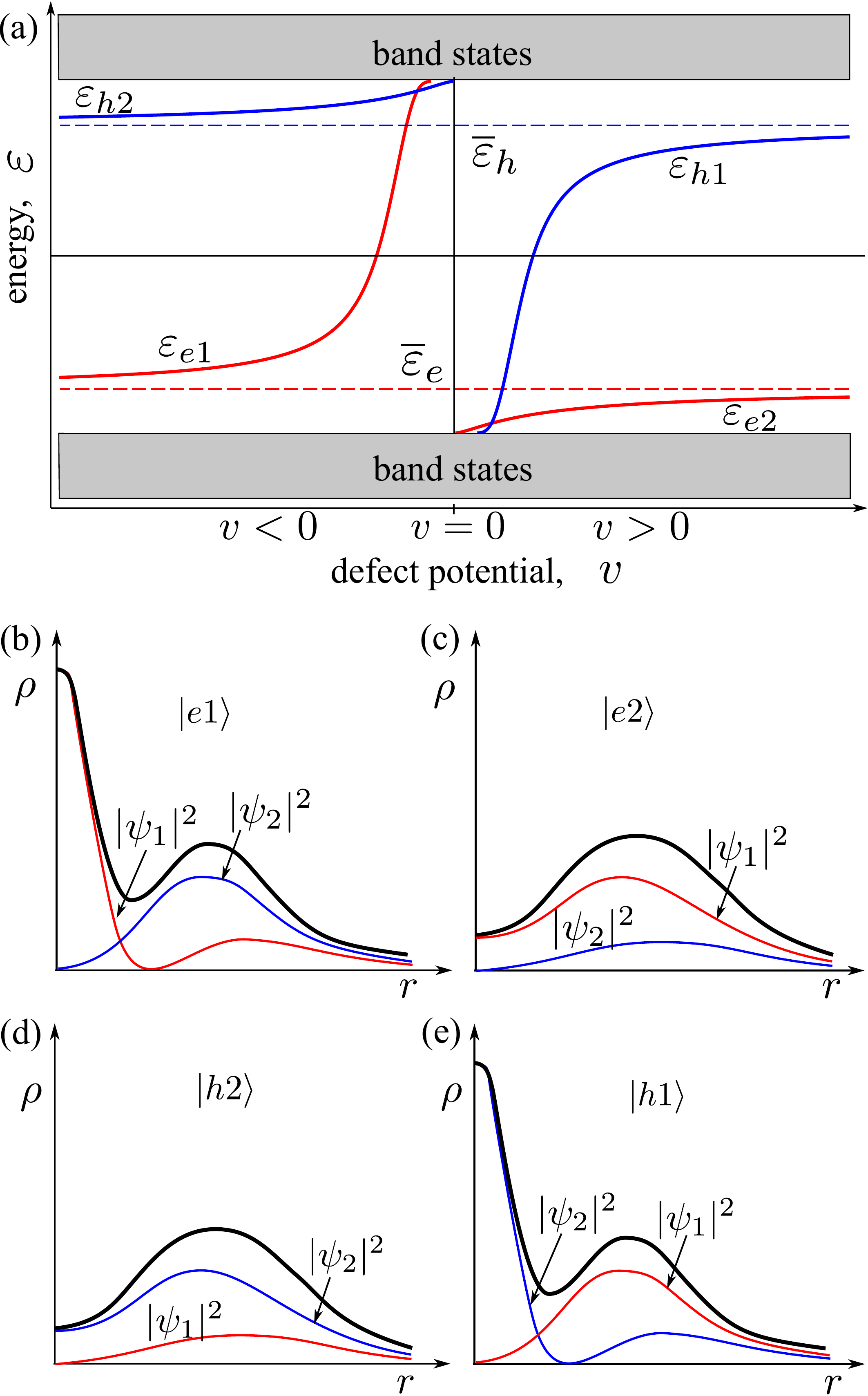}}
\caption{(Color online) A schematic view of (a) the energy of the bound states in the bulk of 2D TI as a function of the defect potential, (b--e) the radial distribution of the electron density $\rho$ (thick lines) and the densities of the spinor components $|\psi_{1,2}|^2$ for electronlike and holelike bound states of the first and second types.}
\label{f_b-st_bulk}
\end{figure}

Unusual properties of the bound states in 2D TIs become apparent from the dependence of the bound state energies on the defect potential amplitude $v$. They are illustrated in Fig.~\ref{f_b-st_bulk}(a). The energies of both electronlike and holelike states have two branches with the quite different dependence of the energy on $v$. To be specific, we consider the electronlike states. One branch, $\varepsilon_{e1}(v)$, appears when the potential is attractive for electrons, $v<0$. As $|v|$ increases, the bound state $|e1\rangle$ appears with the energy at the top of the gap. Thereafter, its energy goes to the bottom of the gap, reaching asymptotically a limiting value $\overline{\varepsilon}_{e}$. We call such states the states of the first type.

When the potential is repulsive for electrons, there is another branch $\varepsilon_{e2}(v)$, which represents the bound states of the second type, $|e2\rangle$. With increasing $v$, the energy $\varepsilon_{e2}$ changes from the bottom of the gap to the limiting energy $\overline{\varepsilon}_{e}$. The holelike states $|h1\rangle$ and $|h2\rangle$ behave symmetrically with respect to the electron-like states in accordance with Eq.~(\ref{e-h_symmetry}): $\varepsilon_{h(1,2)}(v)=-\varepsilon_{e(1,2)}(-v)$.

A physical difference between the states of the first and second types is seen from the spatial distribution of the electron density $\rho=\Psi^{\dag}\Psi$ and the pseudospin components of the density $|\psi_1|^2$ and $|\psi_2|^2$. Graphs of the radial distribution of the total electron density and the pseudospin components are shown in Figs.~\ref{f_b-st_bulk}(b)--\ref{f_b-st_bulk}(e). In the states of the first type the density $\rho$ has a maximum in the point of the defect location, while in the second-type states the density reaches a minimum at the defect. Nevertheless, one should note that in the general case the maximum of the density in the center is not necessarily the highest maximum in the radial distribution of the density. Under certain conditions, another maximum may appear at some distance from the center.

It is remarkable that in the first-type states, the pseudospin component (electronlike or holelike one) which reaches a maximum in the center is exactly  that for which the defect potential is attractive. The opposite situation occurs for the states of the second type. The component of the spinor which vanishes in the center is that for which the potential is repulsive.

The existence of two states in a short-range potential is a feature of the 2D TIs. In the topologically trivial case, where $MB<0$, the calculations carried out by the same method show that there are also electronlike and holelike states, but only one state arises in a given potential. The electronlike state exists only at $v>0$ and the holelike state exists at $v<0$. Moreover, the states occur in a finite range of $|v|$. In both states, the density $\Psi^{\dag}\Psi$ reaches the maximum in the center, i.e. both states are the states of the first type in our classification. The second-type states are absent.

These facts allow one to interpret the presence of two bound states in a given potential as a result of a simultaneous action of two mechanisms of bound state formation. The first mechanism is universal: the bound states can be formed by the potential attracting the quasiparticles of one of the bands. Another mechanism is specific for TIs. It is caused by the formation of an edge state circulating around the defect similarly to the edge states near the boundary. In a certain sense, the defect effectively creates a boundary condition for the wave function. This mechanism was discussed in the literature~\cite{Shan,J_Lu,Shen}.

The existence of two states agrees qualitatively with recent numerical calculations with using a tight-binding approach combined with the Green's function method~\cite{Lee}, but there are essential discrepancies. The contradictions are clearly seen in the spatial distributions of the total density $\rho$ and the pseudospin components of the densities, as well as in the dependence of the densities on the impurity potential. The results we have obtained here very well agree with the direct calculations within the continuous model of the isolated defect~\cite{Sablikov}.

The bound state energies depend also on the parameter $a$, defined through the parameters of the BHZ model by Eq.~(\ref{dim_less_param}). When $a<2^{1/2}$, the energy gap is less than $|M|$. In this case, two bound states with the energy within this reduced gap exist for all values of the defect potential and the graphs of $\varepsilon_{(e,h)(1,2)}(v)$ look as in Fig.~\ref{f_b-st_bulk}(a). In the case where $a>2^{1/2}$, a qualitative difference arises for the energy of the first-type states. These states appear when $|v|$ exceeds a threshold value. Below the threshold, only the second-type states exist.  

The approach we use, does not allow us to investigate the effect of the shape of the defect potential on the bound states. One can only trace how the bound state energies change with the localization radius when it is small. We considered the case where the localization length changes together with the potential amplitude so that the integral of the potential over the area remains constant. It was found that as the localization length decreases, the limiting energies $\overline{\varepsilon}_{e,h}$ shift slowly (logarithmically) to the nearest edges of the gap. In the limiting case of $\Lambda \to \infty$, the potential shape becomes the $\delta$ function and the bound states disappear in agreement with the theory of singular potentials.~\cite{Jackiw}.

The energy of the bound states does not depend on the spin, which means that there are two states with opposite spins. In these states, the electron current circulates around the defect in opposite directions, just as in the edge states. However, in contrast to them, each state can be occupied by one electron with some spin because the capture of another electron is hindered by Coulomb repulsion. The possibility of capturing another electron with opposite spin requires a separate consideration, taking into account the interaction between electrons.

\section{Defect-induced states near the boundary}\label{states_near_boundary}
When the defect is coupled with the boundary, the electronic states are classified as the edge states flowing around the defect and the bound states in the continuum. In this section, we consider the resonances of edge states flowing around the defect and find out the conditions under which the bound state arises in the continuum of the edge states.

The specific calculations are performed for the $\delta$-like potential of the defect in the form $v(x,y-b)=(v/\pi)\delta\left[x^2+(y-b)^2\right]$ with using a cutoff at $\Lambda$ in the momentum space. This simplification allows us to calculate analytically the integrals over $p$ in Eqs.~(\ref{Psi_Phi_Psi-bar}) and (\ref{K}). The subsequent integration over $k$ is done numerically. 

It is natural to expect that when the defect is located far from the boundary, the mixing of the bound state and the edge states leads to the formation of a resonance with the energy close to the bound-state energy. At a finite distance $b$ between the defect and the boundary, the resonance broadens and the resonant energy deviates from the bound-state energy. It is this mixture of the states that forms the edge states flowing around the defect. They are exactly described by Eqs~(\ref{Psi_Phi_Psi-bar}), (\ref{Phi-Psi_1}), and (\ref{Psi-bar_flowing}). 

Our analysis shows that the resonant energy is very close to the energy $\varepsilon_0$ defined by the roots of Eq.~(\ref{epsilon_0}). In the limit $b\to \infty$, these roots describe the bulk bound states. In the case of arbitrary distance $b$, it becomes essential that the roots are functions of three variables: the defect potential $v$, the distance $b$, and the material parameter $a$. Particularly, the dependence of $\varepsilon_0(v,b,a)$ on $v$ is much more complicated than in the case of the bulk position of the defect. In the following, we consider the dependence of the resonant energy on all three quantities.     

\begin{figure}
\centerline{\includegraphics[width=1.\linewidth]{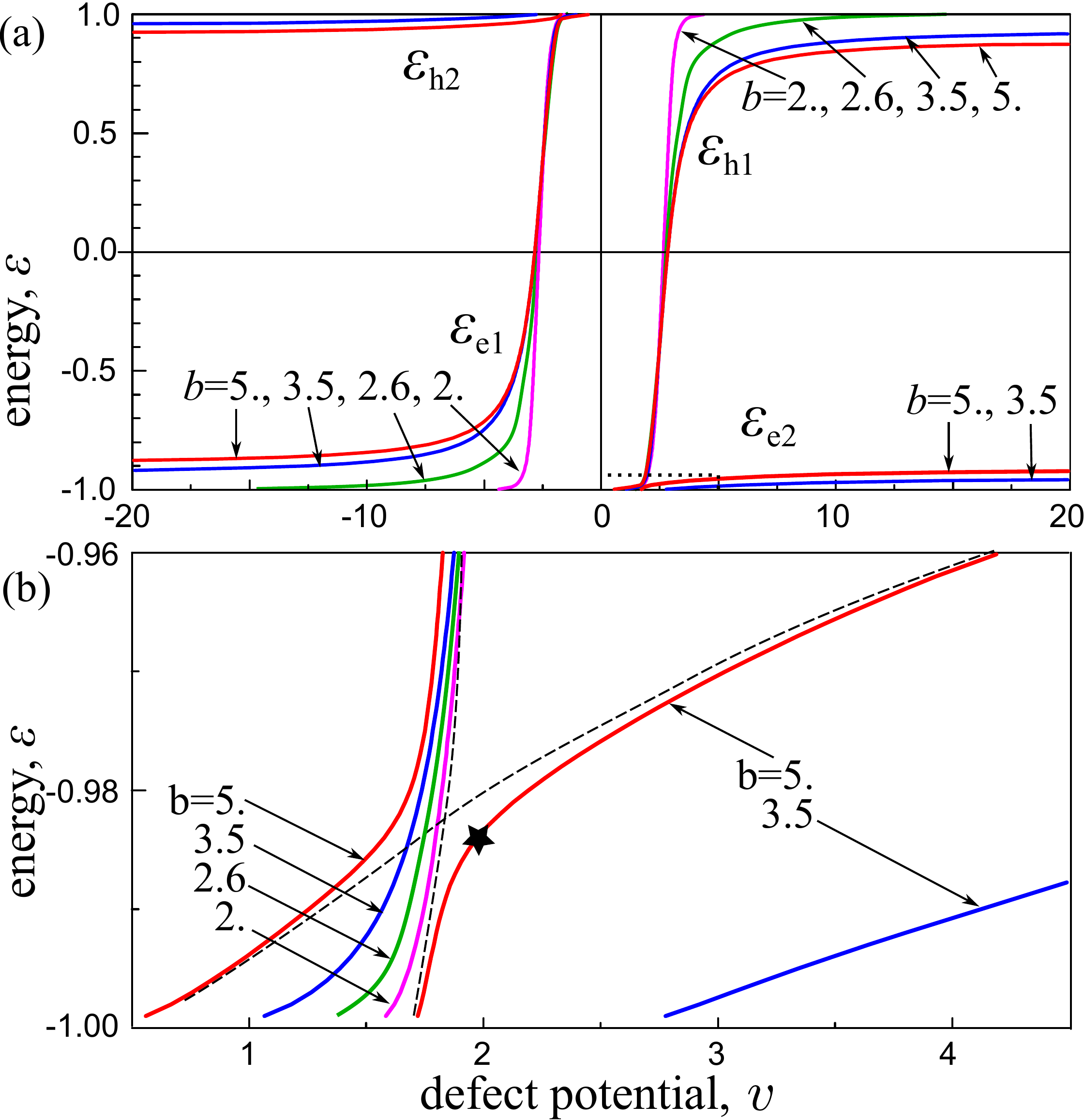}}
\caption{(Color online) (a) Resonant energies $\varepsilon_{(e,h)1}(v)$ and $\varepsilon_{(h,e)2}(v)$ as functions of the defect potential for a variety of distances between the defect and the boundary, $b=2, 2.6, 3.5, 5$. The dotted line shows the region magnified in the lower part of the figure. (b) Graphs $\varepsilon_{(e,h)1}(v)$ and $\varepsilon_{(h,e)2}(v)$ near the anticrossing point. Dashed lines show the crossing curves $\varepsilon_{h1}(v)$ and $\varepsilon_{e2}(v)$ for the bulk states, $b=\infty$. The asterisk shows the energy of the bound state $\varepsilon_{bs}=-0.984$ which exists at $v_{bs}=1.959$ when the defect is located at the distance $b=5$. The calculations were carried out for $a=2^{1/2}$.}
\label{f_r-st_edge}
\end{figure}

When $a>2^{1/2}$, general patterns of the behavior of $\varepsilon_0(v)$ with decreasing $b$ are as follows [see Fig.~\ref{f_r-st_edge}(a) for illustration]. The limiting energies $\overline{\varepsilon}_{e,h}$ shift from their positions at $b\to \infty$ to the edges of the gap so that at some finite distance $b=b_m$ the limiting energy $\overline{\varepsilon}_e$ crosses the bottom of the gap and $\overline{\varepsilon}_h$ crosses the top of the gap. Since our calculations are not justified for the energy outside the gap, we can only conclude that the roots $\varepsilon_{(e,h)2}$ corresponding to the second-type resonances disappear in the gap when $b<b_m$. Nevertheless, the first-type resonances continue to exist, but in a finite interval of $v$, which diminishes with decreasing $b$.

The minimum distance $b_m$ , above which the resonant energy of the second-type states lies in the gap, depends on the parameter $a$. Numerical calculations show that $b_m$ varies slowly between values 2 and 3 when $|a|\lesssim 1.5$ passing through a minimum at $|a|\simeq 1$. With increasing $|a|$ above 1.5, the distance $b_m$ grows sharply, reaches 15 at $|a| = 2.5$, and continues to grow further.

A special situation arises in the region near the intersection point of the curves $\varepsilon_{(e,h)1}(v)$ and $\varepsilon_{(h,e)2}(v)$ calculated in the limit $b\to \infty$. At finite $b$, the coupling of the bound states with the boundary results in a mixing of the states of the first and second types. This leads finally to an avoided crossing of the energy levels $\varepsilon_{(e,h)1}(v)$ and $\varepsilon_{(h,e)2}(v)$. An example of such an anticrossing is shown in Fig.~\ref{f_r-st_edge}(b) for $b=5$. The anticrossing of the levels $\varepsilon_{(e,h)1}(v)$ and $\varepsilon_{(h,e)2}(v)$ occurs only when the distance $b$ is large enough. The minimum distance $b$, above which the anticrossing occurs, depends upon the parameter $a$. With decreasing $a$ the minimum distance diminishes. 

\begin{figure}
\centerline{\includegraphics[width=1.\linewidth]{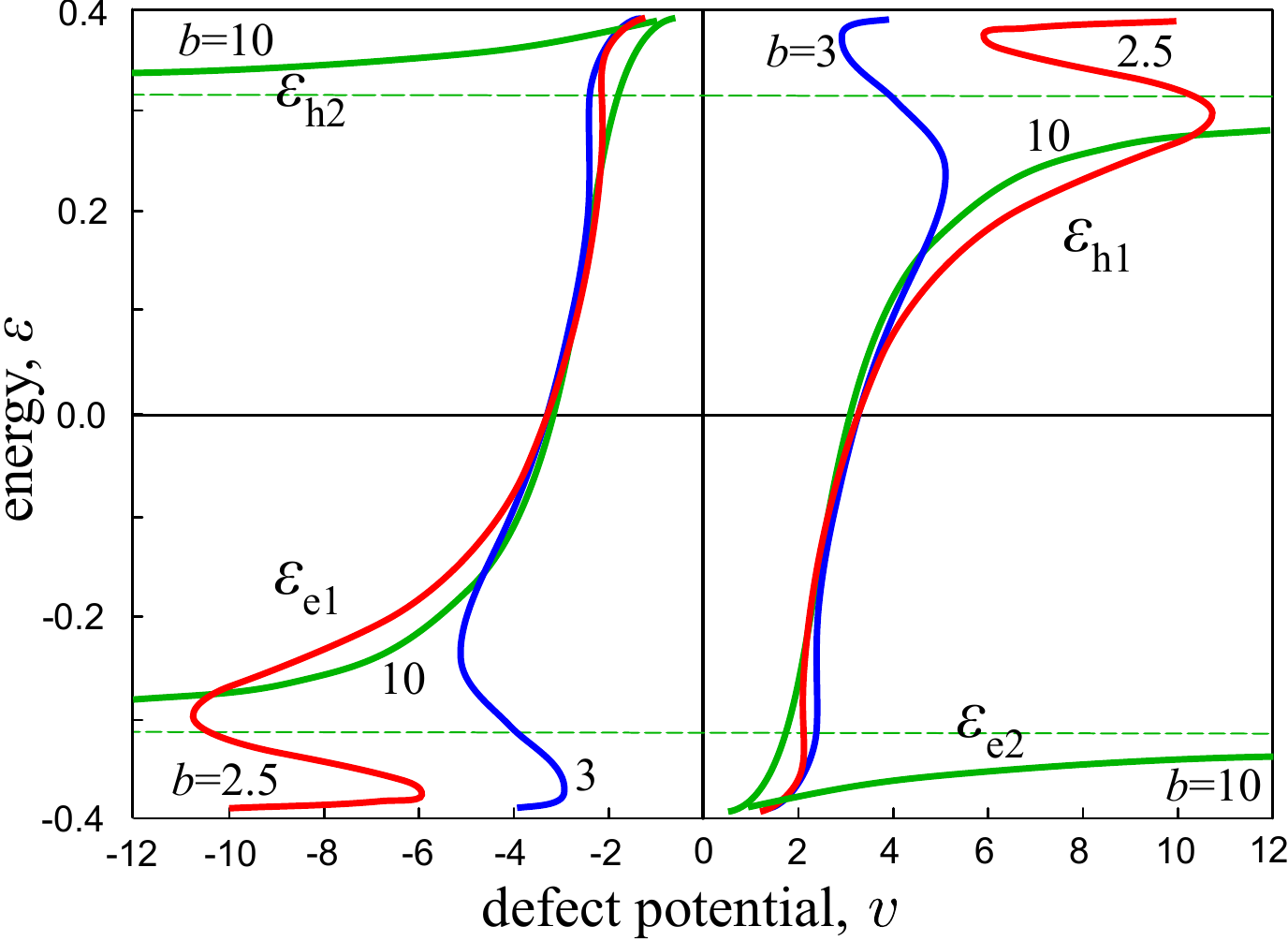}}
\caption{(Color online) The resonant energies $\varepsilon_{(e,h)1}(v)$ and $\varepsilon_{(h,e)2}(v)$ as functions of the defect potential for a variety of the distance between the defect and the boundary, $b=2.5, 3, 10$, in the case where $a=0.4$.}
\label{f_r-st_edge_osc}
\end{figure}

When $a<2$, the wave vectors of the evanescent states, $p_{1,2}(\varepsilon,k)$ defined by Eq.~(\ref{p12}) become complex. However, until the imaginary part of $p_{1,2}$ is small, no new effects appear in the behavior of $\varepsilon_0$ with changing $v$. A qualitatively new behavior appears when $a\ll 1$ and the evanescent states strongly oscillate with distance. Their interference in the segment between the defect and the boundary results in a nonmonotonic dependence of $\varepsilon_0$ on $v$, as it is illustrated in Fig.~\ref{f_r-st_edge_osc}. An oscillatory component of the dependence of $\varepsilon_0$ on $v$ is seen to arise when the defect approaches the boundary. The effect can be so strong that multiple resonant states at a given $v$ can exist under certain conditions. 

These effects could be realized in quantum wells InAs/GaSb where the parameters of the BHZ model are such that $a\sim 0.2\textendash0.5$ and the imaginary part of $p_{1,2}$ greatly exceeds the real part.   

An unexpected effect of the coupling between the defect and the boundary consists in the appearance of the bound state in the continuum of the edge states. In Sec.~\ref{bound_flowing_states}, we have shown that the energy $\varepsilon_{bs}$ of this state and the defect potential $v_{bs}$, at which it appears, are determined by the system of Eqs.~(\ref{epsilon_0}) and (\ref{1-K11-K12}). It is not difficult to analyze these equations in the case of large $b$ where the defect is weakly coupled with the boundary and therefore the non-diagonal elements of the $\mathcal{K}$ matrix given by Eq.~(\ref{K}) are small. In this case, the non-diagonal terms can be treated perturbatively. Finally, we come to the following conclusions: (i) the bound state can arise at those $\varepsilon$ and $v$ which are located on the plane ($\varepsilon, v$) in the vicinity of the intersection point of the curves $\varepsilon_{(e,h)1}(v)$ and $\varepsilon_{(h,e)2}(v)$ calculated for $b\to\infty$. (ii) in this region there is only one solution, which exists if $b$ is large enough.

Thus, the bound state arises at a definite potential $v_{bs}$ when the defect is located at a distance larger than a threshold value. The energy of this state $\varepsilon_{bs}$ and the defect potential $v_{bs}$ are close to point where two resonances are degenerate. The results of specific numerical calculations for $b=5$ and $a=2^{1/2}$ are shown in Fig.~\ref{f_r-st_edge}(b). The bound-state energy is indicated by the asterisk. 

This qualitatively new property of the defect-induced states is caused by the presence of two types of the resonant states in 2D TIs. The bound state in the continuum arises due to the interference of two resonances tuned by changing the defect potential so that they can be driven into degeneracy. This mechanism is consistent with the theory by Friedrich and Wintgen~\cite{Friedrich}.

\section{Electronic structure of states flowing around the defect}\label{e_dens}
In this section, we present the results of our calculations of the spatial distribution of the electron density and current density in the resonant states flowing around the defect.

\begin{figure}
\centerline{\includegraphics[width=1.\linewidth]{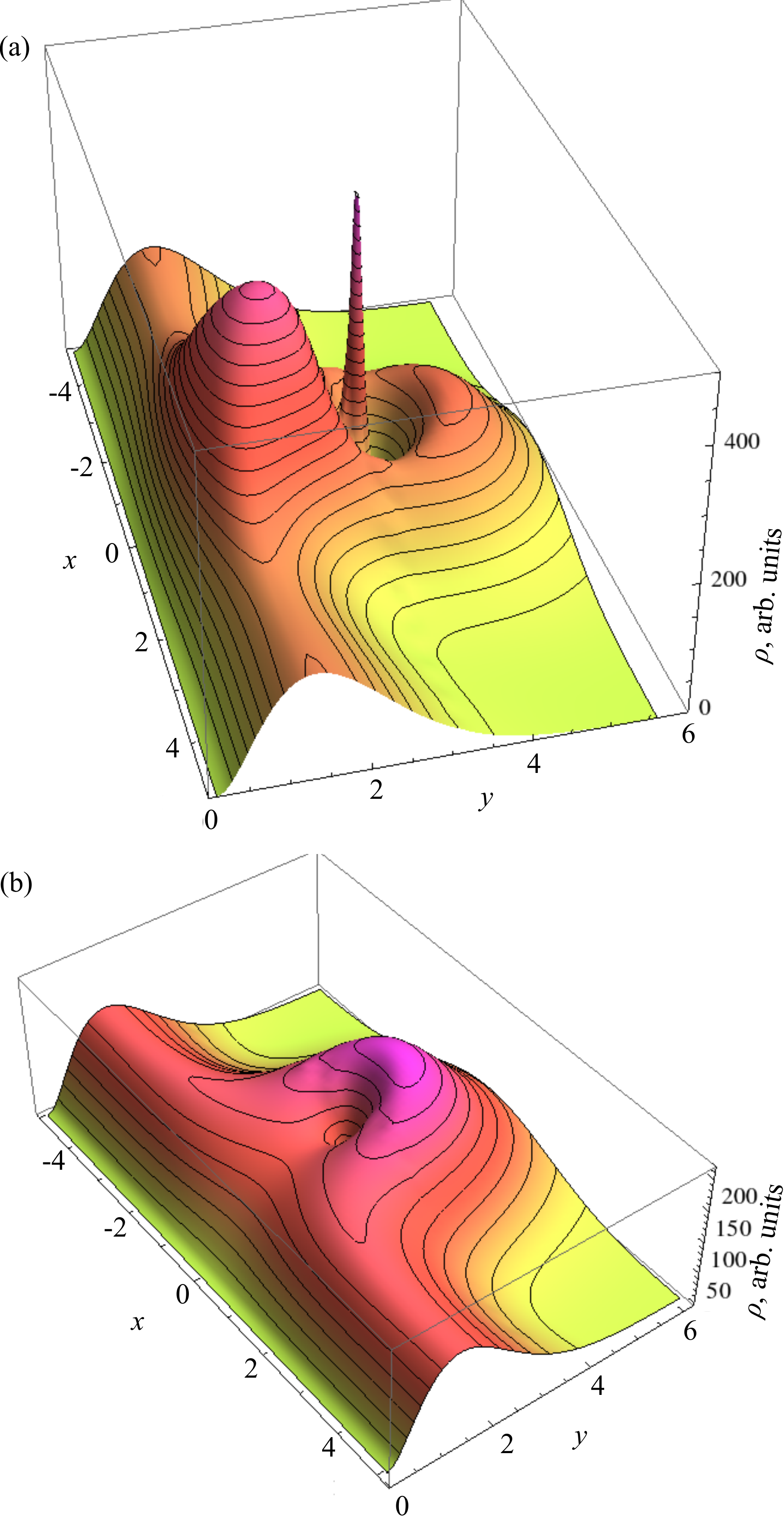}}
\caption{(Color online) 3D plots of electron density distribution in 2D TI with the defect located at the distance $b=3$ from the boundary for two states with different energies close to the resonances of the first and second types: (a) $\varepsilon=0.7$ (first-type resonance) and (b) $\varepsilon=-0.955$ (second-type resonance). Other parameters: $v=3.6$, $a=2^{1/2}$, $\Lambda=10$.}
\label{f_3D_dens}
\end{figure}

The electron density in the edge state flowing around the defect at a given energy, 
\begin{equation}
 \rho_{\varepsilon}(x,y)=\Psi_{\varepsilon}^{\dag}(x,y)\Psi_{\varepsilon}(x,y),                                                                                                                                                                    \end{equation}  
is calculated using $\Psi_{\varepsilon}$ defined by Eq.~(\ref{Psi_Phi_Psi-bar}), where $\Phi(\varepsilon,k)$ and $\overline{\Psi}(\varepsilon)$ are given by Eqs.~(\ref{Phi-Psi_1}) and (\ref{Psi-bar_flowing}). The results obtained for the defect located at the distance $b=3$ from the boundary are presented in Fig.~\ref{f_3D_dens}. The calculations were carried out for the $\delta$-like defect potential. The amplitude of the potential, $v=3.6$, was chosen such that one of the resonances $\varepsilon_{h1}$ was lying deep in the gap, and the other resonance $\varepsilon_{e2}$ was shallow (see Fig.~\ref{f_r-st_edge}). For this potential, we used two energies, one of which $\varepsilon=0.7$ was close to the deep resonance and the other $\varepsilon=-0.955$ was close to the shallow resonance.

Figure~\ref{f_3D_dens} clearly shows that the resonances retain the main properties of the corresponding bound states in the bulk. In the resonances originating from the bound states of the first type, the density has a maximum in the center, while in the second-type resonances the density has a minimum. In contrast to the bound states in the bulk, no component of the spinor $\overline{\Psi}$ equals exactly zero because the resonant states are a mixture of the bulk bound state and the edge states. Nevertheless, one of the spinor components remains much smaller than the other. 

Now, it is interesting to clarify what charge of electrons is accumulated near the defect. The excess electron density at an energy level $\varepsilon$ is evaluated as
\begin{equation}
 \Delta\rho_{\varepsilon}(x,y)=\rho_{\varepsilon}(x,y)-\rho_{\varepsilon}(x,y)\bigm|_{v=0}\,,
\end{equation} 
where the second term in the right hand side is the density in the absence of the defect. To simplify the presentation of the results, we will characterize the excess number of electrons by an integral value $\Delta N(\varepsilon)$ defined as follows:
\begin{equation}
 \Delta N(\varepsilon)=\lim_{L\to\infty}\int\limits_{-L/2}^{L/2}\!dx\!\int\limits_0^{\infty}\!dy\,\Delta\rho_{\varepsilon}(x,y)\,.
\end{equation} 
$\Delta N(\varepsilon)$ gives the spectral density of excess electrons in the states flowing around the defect.

\begin{figure}
\centerline{\includegraphics[width=1.\linewidth]{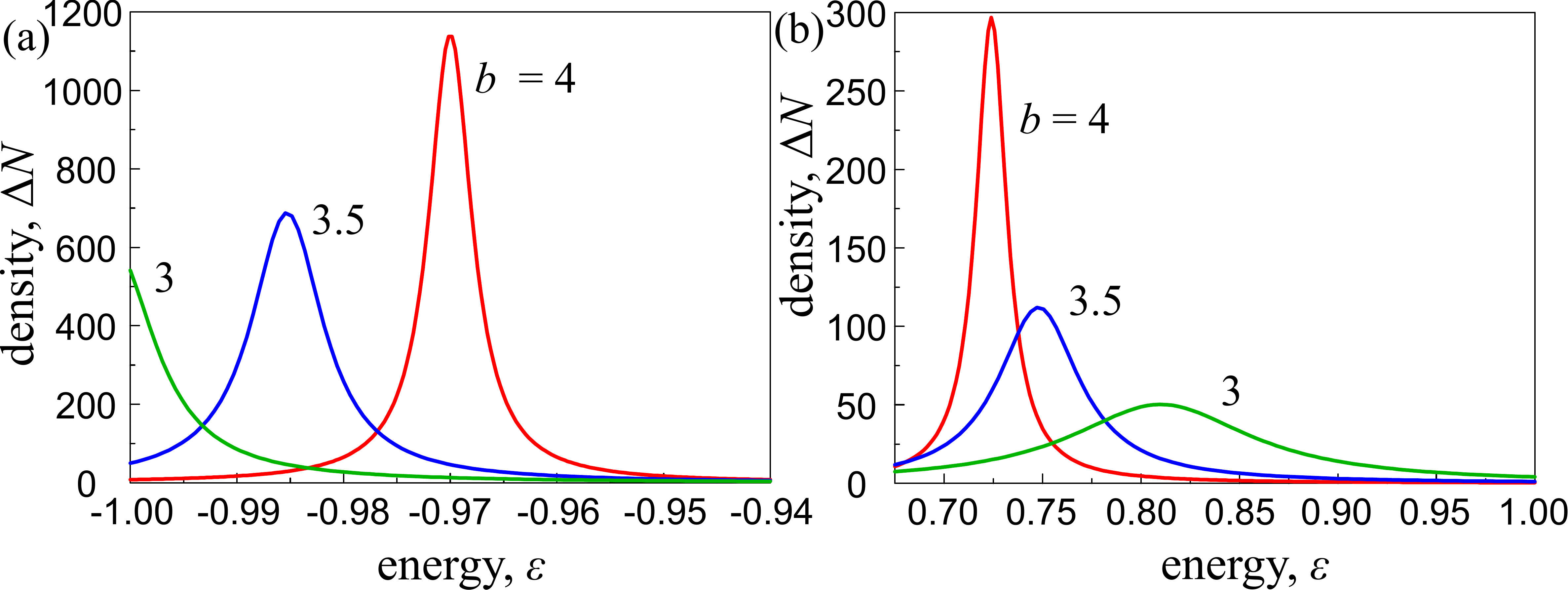}}
\caption{(Color online) Spectral density of excess electrons $\Delta N(\varepsilon)$ as a function of the energy for a variety of $b=$3, 3.5, 4 in two energy ranges: (a) near the resonance of the second type, and (b) near the resonance of the first type. The parameters used in the calculations: $a=2^{1/2}$, $v=5$, and $\Lambda$=10.}
\label{f_N_dens}
\end{figure}

Direct calculations show that $\Delta N(\varepsilon)$ has a maximum at the resonant energies. The results of specific calculations of $\Delta N(\varepsilon)$ are presented in Fig.~\ref{f_N_dens} for different positions of the defect at a given amplitude of the potential. When the defect is located far from the boundary, the shape of the peaks on the curve $\Delta N(\varepsilon)$ is well described by the Breit-Wigner formula. When the distance $b$ is less than the characteristic lengths of the evanescent states, the shape of the peak substantially changes. Nevertheless, the integral of $\Delta N(\varepsilon)$ over $\varepsilon$ within the gap is close to unity if the peak lies far from the gap edges.

This fact allows one to estimate the charge accumulated near the defect. If the Fermi energy is close to the resonant energy, the accumulated charge is of the order of one electron charge. If one assumes that this charge is localized in the area of the radius of about $30$~nm (this is a typical estimation of the decay length of the in-gap states in HdTe quantum wells), the potential created by this charge can be of the order of 10~mV, which is comparable with the gap energy.

The electron flow in the edge states is strongly disturbed in the vicinity of the defect. To clarify the structure of the current field in the presence of a defect one needs to have an explicit expression for the particle current density $\mathbf{j}$ in the quantum states defined by the Hamiltonian of the BHZ model with the defect. Since the Hamiltonian is block diagonal with respect to the spin, it is enough to find the current for one of the spin blocks. The current density is determined by the term of a divergence in the continuity equation. For the spin-up electrons the current $\mathbf{j}_{\uparrow}$ is expressed via the spinor components $\psi_1(x,y)$ and $\psi_2(x,y)$ as follows:
\begin{multline}
 j_{\uparrow x}=-\frac{2i}{\hbar}\mathrm{Im}\left[(B+D)\psi^*_1\frac{\partial \psi_1}{\partial x}-(B-D)\psi^*_2\frac{\partial \psi_2}{\partial x}\right]\\
 +\frac{2A}{\hbar}\mathrm{Re}\left[\psi_1\psi^*_2\right],
\label{jx}
\end{multline} 
\begin{multline}
 j_{\uparrow y}=-\frac{2i}{\hbar}\mathrm{Im}\left[(B+D)\psi^*_1\frac{\partial \psi_1}{\partial y}-(B-D)\psi^*_2\frac{\partial \psi_2}{\partial y}\right]\\
 +\frac{2A}{\hbar}\mathrm{Im}\left[\psi_1\psi^*_2\right].
 \label{jy}
\end{multline} 

When the defect is located in the bulk, a circular electron current is present in each bound state with a given spin~\cite{Sablikov}. Its direction is locked to the spin as in the edge states. However, there is an essential difference from the edge states. The edge state can be occupied by two counter-moving electrons with opposite spins so that the total current of the filled state is zero. In contrast, in the case of a point defect, the problem of the electronic structure of the bound state with two electrons, and even its very existence, requires a separate study taking into account the electron-electron interaction. This issue is beyond the scope of this paper. Within the present approach we study the one-electron states. If one makes the natural assumption that the two-electron state has a higher energy than the one-electron state, then the one-electron bound state is realized. In this state, there is the electron current, whose direction depends on the spin of the trapped electron.   

\begin{figure}
\centerline{\includegraphics[width=1.\linewidth]{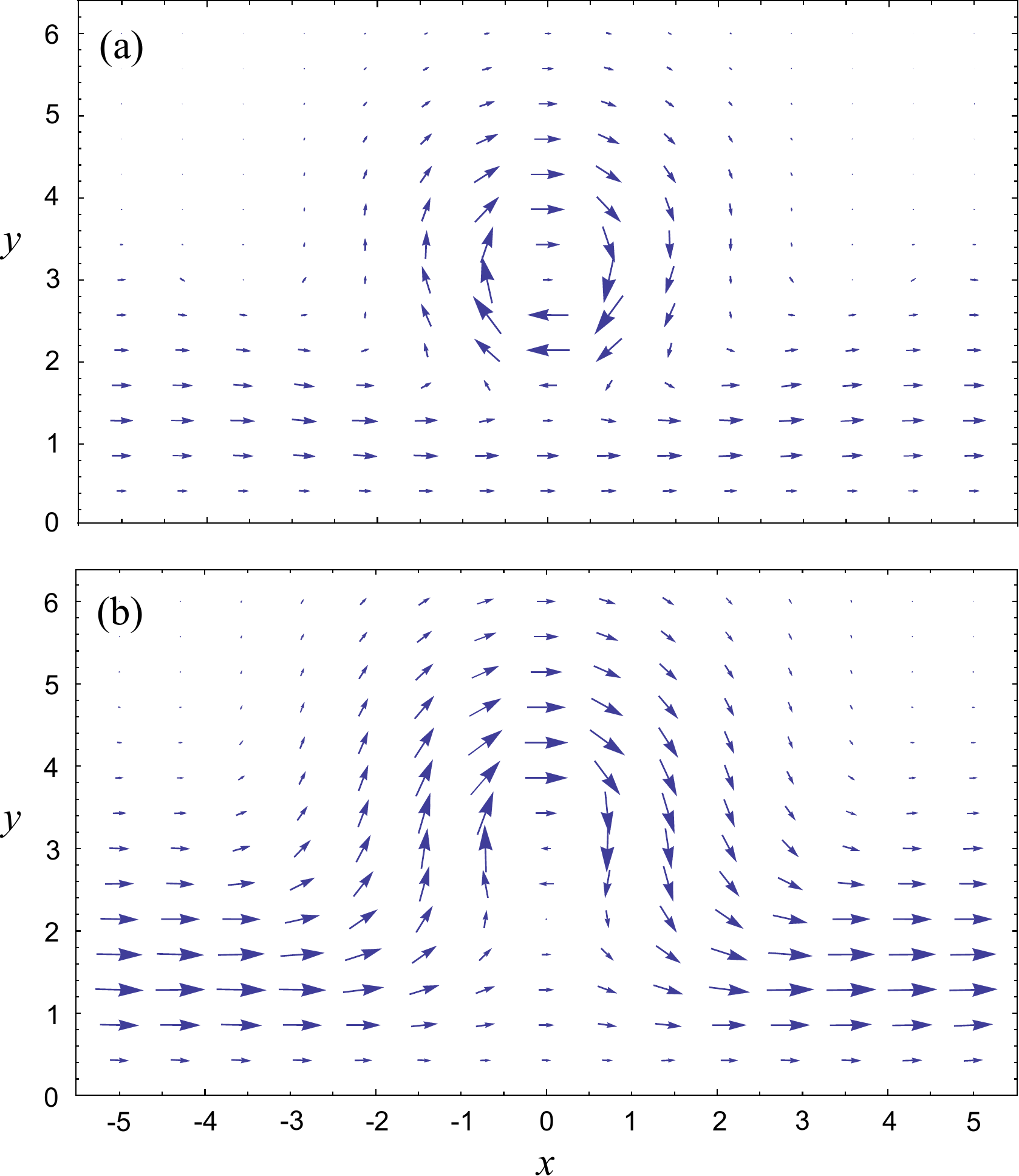}}
\caption{(Color online) Vector plot of the electron current flow of spin-up electrons in the presence of the defect located at the distance $b=3$ from the boundary for two states with energy (a) $\varepsilon$=0.7 near the resonance of the first type and (b) $\varepsilon$=-0.955 near the resonance of the second type. Other parameters are $v=3.6$, $a=2^{1/2}$, $\Lambda$=10.}
\label{f_j}
\end{figure}

If the defect is coupled with the boundary, the field of the current density in the edge state flowing around the defect includes both the current circulating around the defect and the edge current. The configuration of the current field is calculated using Eqs.~(\ref{jx}) and (\ref{jy}). The results are presented in Fig.~\ref{f_j} for two states with the energies lying near the resonances of the first and second types. In both cases there is a current circulating around the defect and an edge current flowing around it. As the defect approaches the boundary, the circulating current decreases and the current flowing around the defect increases. 

\section{Summary and concluding remarks}\label{discuss}

We have studied the in-gap electronic states induced by a nonmagnetic defect with short-range potential in 2D TIs and trace their evolution as the distance between the defect and the boundary changes.

If the defect is located far from the boundary in the bulk, there are two bound states localized at the defect. They exist for both positive and negative potentials of the defect. The states are classified as electron-like and hole-like states depending on the pseudospin (orbital) component of the wave function which vanishes at the center: in the electron-like states $\overline{\psi}_1\ne 0$ and $\overline{\psi}_2=0$ while in hole-like states $\overline{\psi}_1=0$ and $\overline{\psi}_2\ne 0$. In their turn these states are classified into two types depending on the spatial distribution of the particle density and the densities of the pseudospin components. In the states of the first type the particle density has a maximum in the center while the second-type states are characterized by the presence of a minimum of the density in the center. In the states of both types there is a particle current circulating around the defect. Its direction is locked to the electron spin.   

The presence of two bound states of different types at a given defect potential is a feature of 2D TIs. In topologically trivial insulators, there is only one state. The existence of two states raises an interesting question about the two-particle bound state.

Another interesting property of the bound states in the bulk of 2D TIs is the singular dependence of their energy on the defect potential amplitude $|v|$. The energies of the electronlike and holelike states tend to the corresponding limiting values $\overline{\varepsilon}_e$ and $\overline{\varepsilon}_h$ as $|v|\to \infty$. This fact could lead to a non-trivial consequence in the case where the crystal contains many different defects with potentials that are scattered over a wide range. Since the energy of the strong defects slowly changes with their potential, the bound state energies are concentrated in narrow spectral bands near $\overline{\varepsilon}_e$ and $\overline{\varepsilon}_h$. The states may overlap and form a hopping system, which could manifest itself in the transport.

When the defect is coupled with the boundary, the edge states of the host crystal and the bound states transform into a unique set of the edge states flowing around the defect. These states have two resonances corresponding to the two types of the bound states in the bulk. The resonances retain distinctive properties of the bulk bound states. Particularly, they are classified as the resonances of the first and second types. The states flowing around the defect with the energy close to the first-type resonance have a maximum of the electron density at the defect location point, while in the states with the energy near the second-type resonance the density reaches a minimum at the defect.

When electrons occupy the edge states flowing around the defect, the excess charge is accumulated in the vicinity of the defect. Its magnitude can be as large as one electron charge. The potential created by such a charge can be comparable with the energy gap in 2D TIs. This estimation shows that the electron-electron interaction can essentially modify the defect-induced states.

In the edge states flowing around the defect there is an electron current for each spin orientation. The current density field has two components one of which circulates near the defect and another flows around it. The total current flowing around the defect depends on the filling of the states. In the case of a narrow resonance, even a small difference in the population of the states with opposite spins can lead to a noticeable current in the loop around the defect and the formation of a magnetic moment. Estimations show that the magnetic moment can be as large as one Bohr magneton.

An interesting feature of the bulk bound states is the presence of a point where the bound states of different types are degenerate. This happens at a certain potential of the defect. The degeneracy of the resonances is lifted due to the interaction of the defect with the boundary. The interference of the resonances results in the formation of a bound state embedded into the continuum of the edge states flowing around the defect.

\acknowledgments
This work was partially supported by Russian Foundation for Basic Research (Grant No~14-02-00237) and Russian Academy of Sciences.

\end{document}